\documentclass[]{revtex4}
\usepackage{hyperref}
\begin{document}

\title{The conformal gauge to the  derivative gauge for worldsheet gravity}

 \author{ Sudhaker Upadhyay}
 \email {  sudhakerupadhyay@gmail.com; 
 sudhaker@iitk.ac.in}

\affiliation { Department of Physics, Indian Institute of Technology Kanpur, Kanpur 208016, India}

\begin{abstract}
  The BRST quantizations of worldsheet gravity corresponding to final more acceptable derivative gauge and   the standard conformal gauge  are studied. We establish a mapping between
  these two gauges utilizing FFBRST formulation in standard way.
  Therefore, we are able to declare that the problems associated with Virasoro constraints 
  are the gauge artifact. 
 \end{abstract}

\maketitle 
\section{Introduction} 
It has been found that the BRST formalism  is helpful in deriving the full spectrum of low-dimensional string and W-string theories \cite{bl,sk}. For instance, in the handling of anomalies in world-sheet chiral algebras the appropriateness of the BRST
formalism gives full control.
 In case of the non-critical bosonic string,  the presence of a propagating Liouville mode which is originated by the worldsheet anomaly makes the worldsheet “gravity” non-trivial.
A  worldsheet $W_3$ gravity described by an
$A_2$ Toda theory is produced by anomalies in the $W_3$  string \cite{sr}. 
The anomalous Ward identities description for nonlinear chiral worldsheet algebras such
as $W_3$ is made more difficult by the complexity and off-diagonal nature of the anomalies.
The approach made  in \cite{poly, eb, cn}  to the $W_N$ gravity case ran into the difficulty
that a consistent set of conditions to impose on the background gauge fields to eliminate the
anomalies could not be derived owing to their off-diagonal structure  was extended   in \cite{ks}. These difficulties were
actually related to our incomplete knowledge  of $W_3$ geometry.
Further, a reformulation of the BRST quantization  procedure for worldsheet
gravity and the derivation of anomalous Ward identities were made in \cite{rm} which 
are useful for understanding the dynamics of non-critical worldsheet gravity. 

On the other hand, the BRST formalism has proven to be the most powerful approach to the quantization 
of string/gauge theories.
The  generalization of BRST symmetry, known as finite field-dependent BRST (FFBRST) transformation,  has been studied  firstly in \cite{sdj}. 
Further it has been found enormous   applications in the diverse gauge  theories 
\cite{sb,sdj,sdj1, smm,fs,sud1,susk, subp1,ssb,sudd,sud0,sud001,sud21,sud31,sud01,rbs,rs,fsm}. 
For instance, more recently, 
 the gauge-fixing and ghost terms corresponding to Landau and maximal Abelian gauge have  been   
produced for the  
  Cho-Faddeev-Niemi  decomposed $SU(2)$ theory  using FFBRST transformation
  \cite{sud0}. However, the connection between linear and non-linear gauges for perturbative quantum gravity  at both classical and quantum level has been established utilizing FFBRST transformation \cite{sud001}.
 In another problem, the  quantum gauge freedom studied by gaugeon formalism has also been  addressed
  for quantum gravity \cite{sud21} as well as for Higgs model \cite{sud31}.
The FFBRST transformations have been employed for the lattice gauge theory \cite{rs} and the
relativistic point particle model \cite{rbs}.  
Recently, the such transformation is studied in relatively different manner in \cite{lav,alex}.
However, such formulation has not been discussed so-far
for the  worldsheet gravity. 
This gives us a glaring omission to study such transformation in the context of
Virasoro gravity theory where one needs to fix gauge twice.

In this work, we first develop the methodology for 
FFBRST transformation for the gravity theory as a gauge theory.
In this context we compute the finite Jacobian for the functional measure which
depends on the finite field-dependent parameter implicitly.
Such Jacobian actually modifies the effective action of the theory.
We discuss the BRST quantization  
Virasoro worldsheet gravity from the different gauge perspectives. In this scenario  
 we found that the derivative gauge is actually more acceptable than the standard 
conventional gauge.
Further, we generalize the BRST transformation corresponding to the conventional gauge 
by making the infinitesimal parameter finite and field dependent.
Further, we construct an specific parameter such that the Jacobian corresponding to the path 
integral measure 
takes the theory from  the conventional gauge to the
derivative gauge for worldsheet gravity. Since the problems associated with  Virasoro constraints appear only in conventional gauge but not in the derivative gauge \cite{rm}.
Therefore, we overcome this difficulty by connecting the conventional gauge to the
derivative gauge.

We organize this paper in following way. In section II, we provide the
details of FFBRST mechanism. In section III, we sketch briefly the 
BRST quantization for Virasoro gravity with the help of two examples.
In section IV, we derive FFBRST transformation for such gravity theory to 
establish the connection between the conventional and derivative gauge.
In the last section we summarize the results.
\section{FFBRST transformation: Methodology}
To analyse the FFBRST transformation, we start with the usual BRST transformation  for the 
(generic) fields $\phi$ written compactly  as
\begin{eqnarray}
\delta_b \phi  =s_b \phi \  \eta,
\end{eqnarray}
where $s_b \phi$ is the BRST (Slavnov) variation of the fields and  $\eta$
is an  infinitesimal, anticommuting and
global parameter.  Such transformation is  nilpotent in nature, i.e. $\delta_b^2 =0$, with and/or without use of 
 equation of motion of the antighost fields  called as on-shell and/or off-shell nilpotent respectively. 
  It may be observed  that   to be symmetry of Faddeev-Popov effective action it is not necessary to $\eta$ to be infinitesimal
and field-independent as long as it does not depend on the space-time
explicitly.
In fact the following 
finite  field-dependent BRST (FFBRST)
transformation  has been introduced 
which preserves the same form as the BRST transformation   
\begin{eqnarray}
\delta_b \phi =s_b \phi \ \Theta [\phi ],
\end{eqnarray}
except the field-dependent  parameter $\Theta [\phi ]$ which does not depend on spacetime.

Now, we briefly sketch the necessary steps to construct the FFBRST transformation. 
The first step is to make all the fields $\phi$,  a parameter ($\kappa: 0\le \kappa \le 1$) dependent by continuous interpolation    in 
such a way that fields $\phi(x, \kappa =0 )= \phi(x) $ are the initial fields and
 $ \phi(x, \kappa=1) = \phi ^\prime
(x)$ are the transformed fields. 
Furthermore, the  infinitesimal parameter   $\eta$  is made field dependent which characterizes  the following infinitesimal field-dependent BRST transformation:
\begin{equation}
\frac{ d}{d \kappa}\phi(x, \kappa ) = s_b \phi (x, \kappa )\
\Theta_b^\prime [\phi(x,\kappa )].
\label{ibr}
\end{equation}
Here $\Theta ^\prime $  denotes the infinitesimal field-dependent parameter.  The  integration of such transformation
from $ \kappa=0$ to $\kappa=1$ leads to the following FFBRST transformations \cite{sdj}
\begin{equation}
\delta_b  \phi(x)=\phi^\prime (x) -\phi(x) = s_b \phi (x)\ \Theta  [\phi],
\label{fbrs}
\end{equation}
where $\Theta  [\phi] $ is (an arbitrary) finite field dependent parameter. The parameters  $\Theta  [\phi] $ and  $\Theta ^\prime [\phi]$ are related by 
  \cite{sdj} 
\begin{equation}
\Theta  [\phi(x, \kappa)] = \Theta ^\prime [\phi(x)] \frac{ \exp f[\phi(x)]
-1}{f[\phi(x)]},
\label{80}
\end{equation}
where the functional $f[\phi]$ is given by 
\begin{equation}
 f[\phi]= \sum_{i} \frac{ \delta \Theta ^\prime (x)}{\delta
\phi_i(x)} s_b \phi_i(x).
\end{equation}  
The resulting FFBRST transformation leaves the Faddeev-Popov  effective action invariant.
However the path integral measure defined by (${\cal D}\phi$) and therefore the
generating  (vacuum to vacuum) functional
 defined by
\begin{equation}
Z[0]=\int [{\cal D}\phi ]\ e^{iI},\label{generator}
\end{equation} 
get changed non-trivially under such FFBRST transformation. Therefore, the Jacobian is responsible for these changes.
 Now to compute the Jacobian for path integral measure we first write  
\begin{eqnarray}
{\cal D}\phi =J[\phi(\kappa)] {\cal D}\phi(\kappa).  
\end{eqnarray}               
We know that   this non-trivial Jacobian can be replaced (within the 
functional integral) by the local polynomial as \cite{sdj} 
\begin{equation}
J[\phi(\kappa)] \rightarrow e^{iS_1[\phi(\kappa)]},\label{s1}
\end{equation}
where $S_1[\phi(\kappa)]$ is the local functional of fields  $\phi (x)$, iff the following condition gets satisfy:
\begin{eqnarray}
\int [{\cal D}\phi] \left[\frac{1}{J}\frac{d J}{d\kappa} -i\frac{dS_1}{d\kappa}\right] \exp {i[I+S_1]}=0,
\label{mcond}
\end{eqnarray}
where the change in Jacobian has the following explicit expression:
\begin{equation}
 \frac{1}{J(\kappa)}\frac{dJ(\kappa)}{d\kappa}= - \int d^2z\sum_\phi  \left[\pm s_b\phi \frac{\delta \Theta_b'[\phi(\kappa)]}{
\delta\phi(\kappa)}\right].\label{jac}
\end{equation}
Consequently under such process our original generating functional modifies as follows:
\begin{equation}
 \int [{\cal D} \phi]\ e^{i I[\phi]} \stackrel{FFBRST}{----\longrightarrow }\int J[\phi][{\cal D}
\phi] 
\ e^{i \left(I[\phi]\right)}= \int [{\cal D}
\phi] 
\ e^{i \left(I[\phi]+S_1 [\phi]\right)}.
\end{equation} 
Here $S_1[\phi]$ is not an arbitrary functional rather it depends on the choice of finite field-dependent parameter. Therefore,  the two different effective actions can be related through FFBRST transformation with 
appropriate choices of finite parameter. 
\section{BRST Quantisation of Virasoro (W$_3$) gravity}
In this section, we analyse the theory in conventional conformal gauge and the derivative gauge
and their importance.
\subsection{Conventional BRST quantization }

Similar to the bosonic string, that  undergoes a preliminary stage of gauge fixing that
includes the condition in complex light-cone variables $z,\bar z$ of type
\begin{eqnarray}
\gamma_{ij} = \left (\begin{array}{clcr}
          0&1\\
         1&h 
         \end{array}\right),
\end{eqnarray}
 the chiral Virasoro gravity action in the preliminary gauge is defined by
\begin{eqnarray}
I_1=\frac{1}{\pi}\int d^2z\left(-\frac{1}{2}\bar\partial\varphi^i\partial\varphi^i+\frac{1}{2}h\partial\varphi^i\partial\varphi^i     \right),\label{act}
\end{eqnarray}
where   $\varphi^i ( i = 0, 1, . . . ,D -1)$, refers a set of matter
fields and $h$ denotes the remaining unfixed component of the two-dimensional metric.
The action (\ref{act}) is invariant under the following gauge transformation:
\begin{eqnarray}
\delta\varphi^i &=&\varepsilon\partial\varphi^i,\nonumber\\
\delta h&=&\bar{\partial}\varepsilon +\varepsilon\partial h-\partial\varepsilon h,
\end{eqnarray}
where $\varepsilon$ is a bosonic parameter of transformation.
To remove the redundancy in gauge degrees of freedom due to gauge symmetry we choose the 
the final conventional
conformal gauge condition $h=h_{back}$. Incorporating this  at quantum level 
we get the following action:
\begin{eqnarray}
I_1=\frac{1}{\pi}\int d^2z\left(-\frac{1}{2}\bar\partial\varphi^i\partial\varphi^i -b\bar{\partial}c+\pi_h(h-h_{back}) -h(T_{mat} +T_{gh})     \right). \label{act1}
\end{eqnarray}
Here $\pi_h$ is an auxiliary field and $b, c$ are Faddeev-Popov ghost fields. $T_{mat}$ and $T_{gh}$ are the energy-momentum tensors for the matter fields and
ghost fields respectively, having following expressions:
\begin{eqnarray}
T_{mat}&=&-\frac{1}{2}\partial\varphi^i\partial\varphi^i,\nonumber\\
T_{gh}&=& -2b\partial c-\partial bc.
\end{eqnarray}
Now the effective action (\ref{act1}) respects the following BRST symmetry:
\begin{eqnarray}
\delta_b\varphi^i &=&-c\partial\varphi^i\eta,\nonumber\\
\delta_b h&=& -(\bar{\partial}c +c\partial h-\partial c h)\eta,\nonumber\\
\delta_b c&=& c\partial c\eta,\nonumber\\
\delta_b b&=&\pi_h\eta,\nonumber\\
\delta_b \pi_h &=& 0,\label{brst}
\end{eqnarray}
where $\eta$ denotes the anticommuting global parameter.
The physical state can be spanned by restricting it with the help of Noether's charge $Q=\int dz c\left(T_{mat}+\frac{1}{2}T_{gh} \right)$ as follows
$Q|\mbox{phys}\rangle =0$.
\subsection{Derivative gauge BRST quantization}
In this subsection, we fix the final gauge of Virasoro gravity by choosing the derivative gauge condition $\bar\partial h = 0$ 
rather than the conventional gauge.   For  this gauge choice the action  in the preliminary gauge (\ref{act}) gets the 
following expression:
\begin{eqnarray}
I_2=\frac{1}{\pi}\int d^2z \left(  -\frac{1}{2} \bar{\partial}\varphi^i \partial\varphi^i -hT_{mat} +\pi_h\bar{\partial}h -b\bar{\partial}(\bar\partial c+c \partial h-\partial c h)\right).\label{act2}
\end{eqnarray}
Here we note that due to the derivative gauge condition, the ghost action becomes  second order in $\bar{\partial}$
derivatives. To use the canonical formalism, we need to introduce auxiliary fields in order to put the ghost sector into first-order form. Therefore, we   define conjugate momenta corresponding to
the fields $c$ and $b$,
\begin{eqnarray}
\pi_c &=&-\bar\partial b,
\nonumber\\
\pi_b &=& \bar\partial c+c \partial h-\partial c h.
\end{eqnarray}
With the help of these momenta  the second-order action (\ref{act2}) can be written in first-order form as
\begin{eqnarray}
I_2=\frac{1}{\pi}\int d^2z \left(  -\frac{1}{2} \bar{\partial}\varphi^i \partial\varphi^i  +\pi_h\bar{\partial}h -\pi_b \bar\partial b-\pi_c \bar\partial c-\pi_b\pi_c -h(T_{mat}+T_{gh})\right),
\label{act3}
\end{eqnarray}
where the expression of $T_{gh}$ is given by
\begin{eqnarray}
T_{gh}=-2\pi_c \partial c-\partial \pi_c c.
\end{eqnarray}
The effective action (\ref{act3}) remains invariant under following 
BRST transformations:
\begin{eqnarray}
\delta_b\varphi^i &=&-c\partial\varphi^i\eta,\nonumber\\
\delta_b h&=& -\pi_b\eta,\nonumber\\
\delta_b c&=& c\partial c\eta,\nonumber\\
\delta_b \pi_c&=& ( T_{mat}+T_{gh})\eta,\nonumber\\
\delta_b b&=&\pi_h\eta,\nonumber\\
\delta_b \pi_b&=& 0,\nonumber\\
\delta_b \pi_h &=& 0.
\end{eqnarray}
Here these transformations  are now canonical.
 The conserved charge corresponding to such symmetry is calculated using Noether's theorem as
\begin{eqnarray}
Q=\int dz\left( c (T_{mat}+\frac{1}{2}T_{gh})+\pi_h\pi_b\right).
\end{eqnarray}
This charge helps in constructing the physical state from total Hilbert space.
The consequence of derivative gauge is a considerable simplification of the BRST
formulation, the evaluation of anomalies and the expression of Wess-Zumino
consistency conditions (see for details \cite{rm}).
  \section{FFBRST transformation for Virasoro gravity}
  In this section we generalize the BRST  transformation  (\ref{brst})
  to show that the derivative gauge can naturally be derived by operating
  FFBRST operator on generating functional corresponding to conventional gauge.
  In this context, the FFBRST transformation is constructed by
  \begin{eqnarray}
\delta_b\varphi^i &=&-c\partial\varphi^i\Theta[\phi],\nonumber\\
\delta_b h&=& -(\bar{\partial}c +c\partial h-\partial c h)\Theta[\phi],\nonumber\\
\delta_b c&=& c\partial c\Theta[\phi],\nonumber\\
\delta_b b&=&\pi_h\Theta[\phi],\nonumber\\
\delta_b \pi_h &=& 0,
\end{eqnarray}
where $\Theta[\phi]$ is an arbitrary finite field-dependent parameter.
For different choices of such parameter one may produce different scenario.
For instance, we compute the finite parameter obtainable from the 
following infinitesimal parameter:
\begin{eqnarray}
\Theta'[\phi]=-\frac{1}{\pi}\int d^2z b\left(h- h_{back} -\bar \partial h\right).
\end{eqnarray}
Exploiting the expression (\ref{jac}) we calculate the infinitesimal change in Jacobian as follows
\begin{eqnarray}
\frac{1}{J}\frac{dJ}{d\kappa}=\frac{1}{\pi}\int d^2z\left[ b\bar{\partial}c-\pi_h(h-h_{back}) -h( 2b\partial c+\partial b c)  +\pi_h\bar{\partial}h -b\bar{\partial}(\bar\partial c+c \partial h-\partial c h)   \right].\label{jac11}
\end{eqnarray}
Now, to evaluate the finite Jacobian we choose the following expression for local functional $S_1$  as discussed in condition 
(\ref{s1}): 
\begin{eqnarray}
S_1[\phi,\kappa] &=& \int d^2z \left[\xi_1(\kappa) b\bar{\partial}c+\xi_2(\kappa)\pi_h(h-h_{back}) +\xi_3(\kappa)h T_{gh}  \right.\nonumber\\
&+&\left. \xi_4(\kappa)\pi_h\bar{\partial}h +\xi_5(\kappa)b\bar{\partial}(\bar\partial c+c \partial h-\partial c h)  \right].\label{s11}
\end{eqnarray}
The choices for  constant parameters $\xi_i(\kappa), i=1,2,..,5$
are made in such a way that these must vanish at $\kappa=0$.
The condition (\ref{mcond}) in tandem with (\ref{jac11}) and (\ref{s11}) leads
\begin{eqnarray}
\frac{1}{J}\frac{d J}{d\kappa} -i\frac{dS_1}{d\kappa}&=& \int d^2z \left[
(\xi'_1-\frac{1}{\pi})  b\bar{\partial}c+(\xi'_2+\frac{1}{\pi}) \pi_h(h-h_{back}) +(\xi'_3 -\frac{1}{\pi})h T_{gh}  \right.\nonumber\\
&+&\left.( \xi'_4-\frac{1}{\pi}) \pi_h\bar{\partial}h +(\xi'_5+\frac{1}{\pi}) b\bar{\partial}(\bar\partial c+c \partial h-\partial c h)  \right]=0.
\end{eqnarray}
Equating the coefficients of each terms of the above from LHS  to RHS, we get the following 
(exactly solvable) first-order differentiable equations:
\begin{eqnarray}
 \xi'_1-\frac{1}{\pi} =0,\ \ \xi'_2+\frac{1}{\pi} =0,\ \ \xi'_3 -\frac{1}{\pi} =0,\ \ \xi'_4-\frac{1}{\pi} =0,\ \ \xi'_5+\frac{1}{\pi} =0.
\end{eqnarray}
The  solutions  for the above equations are
\begin{eqnarray}
 \xi'_1 = \frac{1}{\pi}\kappa,\ \ \xi'_2 =-\frac{1}{\pi}\kappa,\ \ \xi'_3 =\frac{1}{\pi}\kappa,\ \ 
 \xi'_4 =\frac{1}{\pi}\kappa,\ \ \xi'_5 =-\frac{1}{\pi}\kappa.
\end{eqnarray}
Plugging these identifications  to (\ref{s11})  we get the exact expression for $S_1[\phi,\kappa]$ as
\begin{eqnarray}
S_1[\phi,\kappa] &=& \frac{1}{\pi}\int d^2z \left[ \kappa b\bar{\partial}c -\kappa \pi_h(h-h_{back}) + 
\kappa h T_{gh}  \right.\nonumber\\
&+&\left.  \kappa \pi_h\bar{\partial}h -\kappa b\bar{\partial}(\bar\partial c+c \partial h-\partial c 
h)  \right],
\end{eqnarray}
which vanishes at $\kappa =0$, however, at $\kappa$ it contributes to calculate 
the finite Jacobian  as follows
\begin{eqnarray}
J=e^{iS_1[\phi,1]} &=& \exp \left[\frac{i}{\pi}\int d^2z \left[  b\bar{\partial}c - \pi_h(h-h_{back}) +  
h T_{gh}  \right.\right.\nonumber\\
&+&\left.\left.  \pi_h\bar{\partial}h - b\bar{\partial}(\bar\partial c+c \partial h-\partial c h)  
\right]
\right].
\end{eqnarray}
With this Jacobian our original generating functional changes as follows
\begin{equation}
 \int [{\cal D} \phi]\ e^{i I_1[\phi]} \stackrel{FFBRST}{----\longrightarrow }  \int [{\cal D}
\phi] 
\ e^{i \left(I_1[\phi]+S_1 [\phi]\right)},
\end{equation} 
where
\begin{eqnarray}
I_1+S_1[\phi, 1] &=&\frac{1}{\pi}\int d^2z \left(  -\frac{1}{2} \bar{\partial}\varphi^i \partial\varphi^i -hT_{mat} +\pi_h\bar{\partial}h -b\bar{\partial}(\bar\partial c+c \partial h-\partial c h)\right),\nonumber\\
&=&I_2[\phi],
\end{eqnarray}
which is an effective action for the derivative gauge.
We may note that the Virasoro constraints (putting by hand) come in the picture
 only in conventional gauge (see \cite{rm} for details).
 However, it is shown there that all the
  problems associated with the  Virasoro constraints get resolved naturally in the derivative gauge case.
It means that these problems depend  on the choice of gauges and hence are the gauge artifact.
Remarkably, using standard FFBRST transformation 
one can switch the theory from the standard conformal gauge  to the derivative gauge which is more acceptable also
in the  sense that the evaluation of anomalies and the expression of Wess-Zumino
consistency conditions. 
  
\section{Conclusion}
The   BRST quantization  procedure for chiral worldsheet
gravity by the adoption of a derivative gauge condition, and the introduction of momenta
in order to put the ghost sector of the theory back into first-order form, are well studied  in \cite{rm}.
In the derivative gauge the BRST formalism for worldsheet gravity produces the formalism canonical in the sense that the BRST transformations of all fields now arise as canonical transformations generated by the BRST charge \cite{rm}.

 In this paper we have  provided the basic mechanism of the FFBRST transformation.
 Further, we have discussed the Virasoro gravity from the BRST perspective
 by considering the standard conventional (conformal) gauge and the derivative gauge.
 The derivative gauge has found more important to deal with such theory
 because  in making of the standard conformal gauge one looses the Virasoro constraints as field equations.
 We have generalized the BRST transformation to obtain the FFRBST transformation
 corresponding to
 the conventional gauge. Notably, we have found that the derivative gauge-fixed action (which is more acceptable)
 can be obtained naturally  (within the functional integral) by operating the FFBRST
 transformation on the generating functional for the Virasoro gravity corresponding to
 the conventional gauge. We have shown this result explicitly by calculation.
 So this inspection allows one to
perform the analysis the theory in conventional gauge where the ghost sector are in first-order, however, wherever it finds difficulty in this gauge one can switch the theory
in derivative gauge by applying FFBRST transformation.
It will be interesting to study the worldsheet gravity in the Batalin-Vilkovisky formulation
as there anomalies are present.
Our analysis might be helpful in the complete understating to $W_3$ gravity.


\begin{thebibliography}{99}
\bibitem{bl} B. Lian and G. Zuckerman, Phys. Lett. B 254, 417 (1991);
E. Witten, Nucl. Phys. B 373,  187 (1992);
E. Witten and B. Zwiebach, Nucl. Phys. B 377, 55 (1992).
\bibitem{sk} S. K. Rama, Mod. Phys. Lett. A 6, 3531 (1991);
C. N. Pope, E. Sezgin, K. S. Stelle and X. J. Wang, Phys. Lett. B 299, 247 (1993);
H. Lu, C.N. Pope, X.J. Wang and K.-W. Xu, Class. Quant. Grav. 11, 967  (1994).
\bibitem{sr} S. R. Das, A. Dhar and S. K. Rama, Mod. Phys. Lett. A 6, 3055 (1991);
Int. J. Mod. Phys. A 7, 2295 (1992);
A. Bilal and J.-L. Gervais, Phys. Lett.  B 206, 412 (1988);
Nucl. Phys. B 314, 646 (1989); B 318,  576 (1989). 
\bibitem{poly} A. M. Polyakov, Phys. Lett. B 101, 207 (1981); Mod. Phys. Lett. A 2, 893 (1987).
\bibitem{eb}  E. Bergshoeff, P. S. Howe, C. N. Pope, E. Sezgin, X. Shen and K. S. Stelle,
Nucl. Phys. B 363, 163 (1991).
\bibitem{cn} C. N. Pope, X. Shen, K.-W. Xu and K. Yuan, Nucl. Phys. B 376, 52 (1992).
\bibitem{ks} K. Schoutens, A. Sevrin and P. van Nieuwenhuizen, Nucl. Phys. B 364, 584 (1991);
B 371, 315 (1992). 


\bibitem{rm} R. Mohayaee, C. N. Pope, K. S. Stelle and K.-W. Xu, Nucl. Phys. B 433, 712 (1995).
 \bibitem{sdj} S. D. Joglekar and B. P. Mandal,   {Phys. Rev.}  { {D 51}}, 1919 (1995).
\bibitem{sdj1} S. D. Joglekar and B. P. Mandal, Int. J. Mod. Phys.
A 17, 1279 (2002).
 \bibitem{susk}   S. Upadhyay,   S. K. Rai and B. P. Mandal,  J. Math. Phys.  {52}, {022301} (2011).
\bibitem{sb} S. Upadhyay and B. P. Mandal,    Eur. Phys. J.  {C 72},  2065 (2012); Annals of Physics {  327}, 2885 (2012);  arXiv:1409.1735 [hep-th]. 

 
\bibitem{smm} S. Upadhyay, M. K. Dwivedi and B. P. Mandal, Int. J. Mod. Phys. A 28, 1350033 (2013);  arXiv:1407.2017 [hep-th].
\bibitem{fs} M. Faizal, B. P. Mandal and S. Upadhyay, Phys. Lett. B 721, 159 (2013).
\bibitem{sud1} S. Upadhyay and B. P. Mandal, Eur. Phys. Lett. {  93}, 31001 (2011).

 \bibitem{subp1} S. Upadhyay and B. P. Mandal,     Mod. Phys. Lett.   {A 25}, { 3347} (2010).
\bibitem{ssb}  B. P. Mandal,  S. K. Rai and S.  Upadhyay,
Eur. Phys. Lett. { 92}, {21001} (2010).
\bibitem{sudd} S. Upadhyay and D. Das, Phys. Lett. B 733, 63 (2014).
\bibitem{sud0} S. Upadhyay, Phys. Lett. B 727, 293 (2013).
 \bibitem{sud001} S. Upadhyay, Annals.  Phys. 340, 110  (2014).
  \bibitem{sud21} S. Upadhyay, Annals.  Phys.  344, 290 (2014).
  \bibitem{sud31} S. Upadhyay and B. P. Mandal,  Prog. Theor. Exp. Phys. 053B04  (2014).
\bibitem{sud01} S. Upadhyay, arXiv:1308.0982 [hep-th]; EPL  104, 61001  (2013);  EPL 105, 21001 (2014).
\bibitem{rbs} R. Banerjee, B. Paul and S. Upadhyay,  Phys. Rev. D 88, 065019 (2013).

\bibitem{rs} R. Banerjee and S. Upadhyay, Phys. Lett. B  734, 369 (2014).
 \bibitem{fsm}M. Faizal, S. Upadhyay and B. P. Mandal, Phys. Lett. B 738, 201 (2014).
\bibitem{lav} P. M. Lavrov and O. Lechtenfeld, Phys. Lett. B 725, 382 (2013).
 \bibitem{alex} A. Reshetnyak, arXiv:1312.2092 [hep-th].
\end{thebibliography}
\end{document}